

%
\newbox\leftpage \newdimen\fullhsize \newdimen\hstitle \newdimen\hsbody
\tolerance=1000\hfuzz=2pt
\def\bigans{b }
\message{ big or little (b/l)? }\read-1 to\answ
\ifx\answ\bigans\message{(This will come out unreduced.}
\magnification=1200\baselineskip=18pt plus 2pt minus 1pt
\hsbody=\hsize \hstitle=\hsize 
\else\def\apans{l }\message{ lyman or hepl (l/h) (lowercase!) ? }
\read-1 to \apansw\message{(This will be reduced.}
\let\lr=L
\magnification=1000\baselineskip=16pt plus 2pt minus 1pt
\voffset=-.31truein\vsize=7truein
\hstitle=8truein\hsbody=4.75truein\fullhsize=10truein\hsize=\hsbody
\ifx\apansw\apans\special{ps: landscape}\hoffset=-.59truein
  \else\hoffset=.05truein\fi
\output={\ifnum\pageno=0 
  \shipout\vbox{\hbox to \fullhsize{\hfill\pagebody\hfill}}\advancepageno
  \else
  \almostshipout{\leftline{\vbox{\pagebody\makefootline}}}\advancepageno
  \fi}
\def\almostshipout#1{\if L\lr \count1=1
      \global\setbox\leftpage=#1 \global\let\lr=R
  \else \count1=2
    \shipout\vbox{\ifx\apansw\apans\special{ps: landscape}\fi 
      \hbox to\fullhsize{\box\leftpage\hfil#1}}  \global\let\lr=L\fi}
\fi
%
\catcode`\@=11 
\newcount\yearltd\yearltd=\year\advance\yearltd by -1900

\def\Title#1#2{\nopagenumbers\abstractfont\hsize=\hstitle\rightline{#1}%
\vskip 1in\centerline{\titlefont #2}\abstractfont\vskip .5in\pageno=0}
\def\Date#1{\vfill\leftline{#1}\tenpoint\supereject\global\hsize=\hsbody%
\footline={\hss\tenrm\folio\hss}}
\def\draftmode{\def\draftdate{{\rm preliminary draft:
\number\month/\number\day/\number\yearltd\ \ \hourmin}}%
\headline={\hfil\draftdate}\writelabels\baselineskip=20pt plus 2pt minus 2pt
{\count255=\time\divide\count255 by 60 \xdef\hourmin{\number\count255}
	\multiply\count255 by-60\advance\count255 by\time
   \xdef\hourmin{\hourmin:\ifnum\count255<10 0\fi\the\count255}}}

\def\nolabels{\def\eqnlabel##1{}\def\eqlabel##1{}\def\reflabel##1{}}
\def\writelabels{\def\eqnlabel##1{%
{\escapechar=` \hfill\rlap{\hskip.09in\string##1}}}%
\def\eqlabel##1{{\escapechar=` \rlap{\hskip.09in\string##1}}}%
\def\reflabel##1{\noexpand\llap{\string\string\string##1\hskip.31in}}}
\nolabels
%
\global\newcount\secno \global\secno=0
\global\newcount\meqno \global\meqno=1
\def\newsec#1{\global\advance\secno by1
\xdef\secsym{\the\secno.}\global\meqno=1
\bigbreak\bigskip
\noindent{\bf\the\secno. #1}\par\nobreak\medskip\nobreak}
\xdef\secsym{}
\def\appendix#1#2{\global\meqno=1\xdef\secsym{\hbox{#1.}}\bigbreak\bigskip
\noindent{\bf Appendix #1. #2}\par\nobreak\medskip\nobreak}
%
%
\def\eqnn#1{\xdef #1{(\secsym\the\meqno)}%
\global\advance\meqno by1\eqnlabel#1}
\def\eqna#1{\xdef #1##1{\hbox{$(\secsym\the\meqno##1)$}}%
\global\advance\meqno by1\eqnlabel{#1$\{\}$}}
\def\eqn#1#2{\xdef #1{(\secsym\the\meqno)}\global\advance\meqno by1%
$$#2\eqno#1\eqlabel#1$$}
%
\newskip\footskip\footskip14pt plus 1pt minus 1pt 
\def\f@@t{\baselineskip\footskip\bgroup\aftergroup\@foot\let\next}
\setbox\strutbox=\hbox{\vrule height9.5pt depth4.5pt width0pt}
\global\newcount\ftno \global\ftno=0
\def\foot{\global\advance\ftno by1\footnote{$^{\the\ftno}$}}
%
%
\global\newcount\refno \global\refno=1
\newwrite\rfile
\def\ref{[\the\refno]\nref}
\def\nref#1{\xdef#1{[\the\refno]}\ifnum\refno=1\immediate
\openout\rfile=refs.tmp\fi\global\advance\refno by1\chardef\wfile=\rfile
\immediate\write\rfile{\noexpand\item{#1\ }\reflabel{#1}\pctsign}\findarg}
\def\findarg#1#{\begingroup\obeylines\newlinechar=`\^^M\pass@rg}
{\obeylines\gdef\pass@rg#1{\writ@line\relax #1^^M\hbox{}^^M}%
\gdef\writ@line#1^^M{\expandafter\toks0\expandafter{\striprel@x #1}%
\edef\next{\the\toks0}\ifx\next\em@rk\let\next=\endgroup\else\ifx\next\empty%
\else\immediate\write\wfile{\the\toks0}\fi\let\next=\writ@line\fi\next\relax}}
\def\striprel@x#1{} \def\em@rk{\hbox{}} {\catcode`\%=12\xdef\pctsign{
\def\semi{;\hfil\break}
\def\addref#1{\immediate\write\rfile{\noexpand\item{}#1}} 
\def\listrefs{
\vfill\eject\immediate\closeout\rfile
\baselineskip=12pt\centerline{{\bf References}}\bigskip{\frenchspacing%
\escapechar=` \input refs.tmp\vfill\eject}\nonfrenchspacing}
\def\startrefs#1{\immediate\openout\rfile=refs.tmp\refno=#1}
\def\figures{\centerline{{\bf Figure Captions}}\medskip\parindent=40pt}
\def\fig#1#2{\medskip\item{Figure ~#1:  }#2}
\catcode`\@=12 
%
\ifx\answ\bigans
\font\titlerm=cmr10 scaled\magstep3 \font\titlerms=cmr7 scaled\magstep3
\font\titlermss=cmr5 scaled\magstep3 \font\titlei=cmmi10 scaled\magstep3
\font\titleis=cmmi7 scaled\magstep3 \font\titleiss=cmmi5 scaled\magstep3
\font\titlesy=cmsy10 scaled\magstep3 \font\titlesys=cmsy7 scaled\magstep3
\font\titlesyss=cmsy5 scaled\magstep3 \font\titleit=cmti10 scaled\magstep3
\else
\font\titlerm=cmr10 scaled\magstep4 \font\titlerms=cmr7 scaled\magstep4
\font\titlermss=cmr5 scaled\magstep4 \font\titlei=cmmi10 scaled\magstep4
\font\titleis=cmmi7 scaled\magstep4 \font\titleiss=cmmi5 scaled\magstep4
\font\titlesy=cmsy10 scaled\magstep4 \font\titlesys=cmsy7 scaled\magstep4
\font\titlesyss=cmsy5 scaled\magstep4 \font\titleit=cmti10 scaled\magstep4
\font\absrm=cmr10 scaled\magstep1 \font\absrms=cmr7 scaled\magstep1
\font\absrmss=cmr5 scaled\magstep1 \font\absi=cmmi10 scaled\magstep1
\font\absis=cmmi7 scaled\magstep1 \font\absiss=cmmi5 scaled\magstep1
\font\abssy=cmsy10 scaled\magstep1 \font\abssys=cmsy7 scaled\magstep1
\font\abssyss=cmsy5 scaled\magstep1 \font\absbf=cmbx10 scaled\magstep1
\skewchar\absi='177 \skewchar\absis='177 \skewchar\absiss='177
\skewchar\abssy='60 \skewchar\abssys='60 \skewchar\abssyss='60
\fi
\skewchar\titlei='177 \skewchar\titleis='177 \skewchar\titleiss='177
\skewchar\titlesy='60 \skewchar\titlesys='60 \skewchar\titlesyss='60
\def\titlefont{\def\rm{\fam0\titlerm}
\textfont0=\titlerm \scriptfont0=\titlerms \scriptscriptfont0=\titlermss
\textfont1=\titlei \scriptfont1=\titleis \scriptscriptfont1=\titleiss
\textfont2=\titlesy \scriptfont2=\titlesys \scriptscriptfont2=\titlesyss
\textfont\itfam=\titleit \def\it{\fam\itfam\titleit} \rm}
\ifx\answ\bigans\def\abstractfont{\tenpoint}\else
\def\abstractfont{\def\rm{\fam0\absrm}
\textfont0=\absrm \scriptfont0=\absrms \scriptscriptfont0=\absrmss
\textfont1=\absi \scriptfont1=\absis \scriptscriptfont1=\absiss
\textfont2=\abssy \scriptfont2=\abssys \scriptscriptfont2=\abssyss
\textfont\itfam=\bigit \def\it{\fam\itfam\bigit}
\textfont\bffam=\absbf \def\bf{\fam\bffam\absbf} \rm} \fi
\def\tenpoint{\def\rm{\fam0\tenrm}
\textfont0=\tenrm \scriptfont0=\sevenrm \scriptscriptfont0=\fiverm
\textfont1=\teni  \scriptfont1=\seveni  \scriptscriptfont1=\fivei
\textfont2=\tensy \scriptfont2=\sevensy \scriptscriptfont2=\fivesy
\textfont\itfam=\tenit \def\it{\fam\itfam\tenit}
\textfont\bffam=\tenbf \def\bf{\fam\bffam\tenbf} \rm}
%
%
\def\noblackbox{\overfullrule=0pt}
\hyphenation{anom-aly anom-alies coun-ter-term coun-ter-terms}
\def\inv{^{\raise.15ex\hbox{${\scriptscriptstyle -}$}\kern-.05em 1}}
\def\dup{^{\vphantom{1}}}
\def\Dsl{\,\raise.15ex\hbox{/}\mkern-13.5mu D} 
\def\dsl{\raise.15ex\hbox{/}\kern-.57em\partial}
\def\del{\partial}
\def\Psl{\dsl}
\def\tr{{\rm tr}} \def\Tr{{\rm Tr}}
\font\bigit=cmti10 scaled \magstep1
\def\biglie{\hbox{\bigit\$}} 
\def\lspace{\ifx\answ\bigans{}\else\qquad\fi}
\def\lbspace{\ifx\answ\bigans{}\else\hskip-.2in\fi} 
\def\boxeqn#1{\vcenter{\vbox{\hrule\hbox{\vrule\kern3pt\vbox{\kern3pt
	\hbox{${\displaystyle #1}$}\kern3pt}\kern3pt\vrule}\hrule}}}
\def\mbox#1#2{\vcenter{\hrule \hbox{\vrule height#2in
		\kern#1in \vrule} \hrule}}  
%
\def\CAG{{\cal A/\cal G}}   
\def\CA{{\cal A}} \def\CC{{\cal C}} \def\CF{{\cal F}} \def\CG{{\cal G}}
\def\CL{{\cal L}} \def\CH{{\cal H}} \def\CI{{\cal I}} \def\CU{{\cal U}}
\def\CB{{\cal B}} \def\CR{{\cal R}} \def\CD{{\cal D}} \def\CT{{\cal T}}
\def\e#1{{\rm e}^{^{\textstyle#1}}}
\def\grad#1{\,\nabla\!_{{#1}}\,}
\def\gradgrad#1#2{\,\nabla\!_{{#1}}\nabla\!_{{#2}}\,}
\def\ph{\varphi}
\def\psibar{\overline\psi}
\def\om#1#2{\omega^{#1}{}_{#2}}
\def\vev#1{\langle #1 \rangle}
\def\lform{\hbox{$\sqcup$}\llap{\hbox{$\sqcap$}}}
\def\darr#1{\raise1.5ex\hbox{$\leftrightarrow$}\mkern-16.5mu #1}
\def\lie{\hbox{\it\$}} 
\def\ha{{1\over2}}
\def\half{{\textstyle{1\over2}}} 
\def\roughly#1{\raise.3ex\hbox{$#1$\kern-.75em\lower1ex\hbox{$\sim$}}}

\font\names=cmbx10 scaled\magstep1
\baselineskip=20pt

\Title{PUPT-1436 (1994), hep-ph/9401351}
{\vbox{\centerline
{Constraints and Transport in Electroweak Baryogenesis} }}
\centerline{\bf\names Michael Joyce,}
\centerline{\bf\names Tomislav Prokopec }
\centerline{ and }
\centerline{\bf\names Neil Turok}
\centerline{Joseph Henry Laboratories}
\centerline{Princeton University, Princeton NJ 08544.}
\bigskip
\baselineskip=24pt
\centerline{\bf Abstract}
\baselineskip=12pt
\noindent
In unconstrained  thermal equilibrium a local potential
for total or fermionic hypercharge
does not bias electroweak anomalous processes. We consider two
proposed mechanisms for
electroweak baryogenesis in this light. In `spontaneous' baryogenesis,
which was argued to  apply in the `adiabatic' limit of thick, slow walls,
a non-zero result was obtained by setting globally conserved charges to
be zero {\it locally}. We show that this is a poor
approximation unless the walls are very thick.
For more realistic wall thicknesses the local equilibrium
approached as the wall
velocity $v_w\rightarrow 0$ has zero baryon number
violation and nonzero global charges on the wall.
In the `charge transport'
mechanism, argued to apply to the case of thin fast walls,
calculations of the magnitude of the asymmetry
also involve the same error. In corrected calculations the
local values of global charges should be determined
dynamically rather than fixed locally to zero.

\Date{October 1994 (Revised)}


\baselineskip=24pt

	It is an attractive idea that the observed
excess of baryons over anti-baryons was produced at
the electroweak phase transition. This could happen through
anomalous baryon number violating electroweak processes being biased
during
the dynamical completion of the phase transition, and then
turned off in the broken symmetry phase (for reviews see
\ref\review{N. Turok, in {\it Perspectives in Higgs Physics},
ed. G. Kane, pub. World Scientific, p. 300 (1992). },
\ref\cknrev{A. Cohen, D. Kaplan and A. Nelson, Ann. Rev. Nucl. Part.
Sci. {\bf 43} (1993) 27.}).
Amongst the various
proposed mechanisms  are two suggested
by Cohen, Kaplan and Nelson (CKN) which use an
analysis of chemical potentials to show how this biasing occurs.
In the first of these, the `spontaneous' baryogenesis mechanism
\ref\cknplb{A. Cohen, D.  Kaplan and A.
 Nelson, Phys. Lett. {\bf B263}, 86  (1991).}
it is argued that
there is an induced potential for fermionic hypercharge on
the bubble wall;
in the second, the `charge transport' mechanism
\ref\cknnpb{A. Cohen, D.  Kaplan and A.
 Nelson, Nuc. Phys. {\bf B373}, 453  (1992).},
the CP-violating reflection of particles
by the bubble wall is argued to produce a local density
of hypercharge (or, when the effects of gauge charge screening are
taken into account, of a global charge related to hypercharge
\ref\cknplbb{A. Cohen, D. Kaplan and A.  Nelson, Phys. Lett. {\bf B294}
 (1992) 57. })
in front of the wall in
the unbroken phase. In both cases an analysis of the chemical
potentials for various charges is shown to lead
to a non-zero chemical potential driving  the
anomalous B violating processes in a given
direction. In this letter we
will reconsider this analysis and show that it contains
a conceptual error
which necessitates revision of these mechanisms.
Our comments also apply to  recent work of  Giudice and
Shaposhnikov, in which   they point
out that CKN failed to include the effect of strong sphalerons
\ref\gs{G. F. Giudice and M. Shaposhnikov, Phys. Lett {\bf B326}, 118 (1994).}.

We begin with the case of `spontaneous' baryogenesis.
We restrict ourselves to the following case. We assume, following CKN,
that the effect of the rolling of the Higgs field during the EW
phase transition in a two Higgs doublet
model is to produce a potential for fermionic hypercharge $\phi_{Y_f}(x)$.
We do not address here the issue raised by Dine and Thomas in
\ref\DT{M. Dine and S. Thomas, Santa Cruz preprint 1994,
SCIPP 94/01} about the validity of this treatment. Irrespective of
the resolution of this issue there is a separate and important
point which we make here.

Firstly we recall briefly the  calculation of $B$ violation
in \cknplb\ in which   CKN impose,
through the introduction
of chemical potentials $\mu_A$,
constraints on
a set of exactly conserved quantum numbers $Q_A$ (like
$B-L$) and those
conserved by the particle interactions considered
to be fast
on the timescales of interest (e.g. the number of right handed $b$ quarks).
The procedure
followed  by CKN is to fix all the conserved numbers to zero,
including baryon number $B$, and then compute the chemical potential
which drives $B$ violation , in the background of the potential $\phi_{Y_f}(x)$
for fermionic hypercharge. For simplicity let us just take
$Y$,   $B$ and $B-L$ to be the relevant conserved charges. We find
\eqn\ef{\eqalign{
  Y&\propto 10\phi_{Y_f}+(10+n)\mu_Y +8\mu_{B-L}+2 \mu_B=0 \cr
       B-L&\propto 8\phi_{Y_f} +8\mu_Y+13 \mu_{B-L}+ 4 \mu_B=0 \cr
       B&\propto \phi_{Y_f}+\mu_Y+2 \mu_{B-L}+ 2 \mu_B=0.
}}
using the relation
\eqn\eg{\eqalign{\mu_i=\Sigma_A q^A_i \mu_A}}
and $n_i \propto k_i \mu_i$ for the particles which
we assume massless. $k_i$ is a counting factor which is
one for fermions and two for bosons, $q^A_i$ is the $Q_A$ charge
of species $i$, and $n$ is the number of Higgs doublets in the
theory.
Solving, we find $\dot{B} \propto \mu_{B} \propto \phi_{Y_f} n$, and
we have apparently
a non-zero $B$ violation rate.

	Now consider a system close to thermal equilibrium (in the sense
that the densities of all particle species are close to their equilibrium
values). Let $n_i$ and $\mu_i$ be the densities and chemical potentials
respectively of the particle species labeled $i$ which we take
to be in local thermal equilibrium. Then the rate
at which a particle species $S_j$ approaches its equilibrium value through a
process $\nu_j S_j \leftrightarrow 0 $ (where e.g.
$A \leftrightarrow 2B + C$ has $\nu_A=1, \nu_B=-2, \nu_C=-1$)
involving that species is
\eqn\ea{\eqalign{\dot{n}_i = - \Gamma_e{\delta F \over T} \nu_i =
-{{\Gamma}_e \over {T} }(\Sigma_j
               \nu_j \mu_j) \nu_i\, ,}}
where $\Gamma_e$ is the equilibrium rate for the process at
temperature $T$
 and $\delta F$ is the
change in free energy in the process.
This equation is derived from detailed
balance considerations.
Applying $\ea$ to the electroweak sphaleron processes
$t_Lt_Lb_L\tau_L \leftrightarrow 0$ and $ t_Lb_Lb_L\nu_{\tau}
\leftrightarrow 0$
which change the total baryon number we find
\eqn\eb{\eqalign{\dot{B}=-  {{\Gamma_s} \over {2T} }
       (3\mu_{t_L}+3\mu_{b_L}+\mu_{\tau_L}+\mu_{\nu_{\tau}})\, , }}
where for simplicity we consider only one family of
quarks and leptons, and $\Gamma_{s}$ is the electroweak sphaleron
rate. We shall in this paper ignore the effect of the loop diagram
discussed by Turok and Zadrozny
\ref\NTJTZa{N. Turok and J. Zadrozny, Phys. Rev. Lett. {\bf 65}, 2331 (1990).},
\ref\NTJTZb{N. Turok and J. Zadrozny,
Nuc. Phys. {\bf B 358}, 471 (1991).},
McLerran et. al.
\ref\MSTV{L. McLerran, M. Shaposhnikov, N. Turok and M. Voloshin,
 Phys. Lett. { \bf  256B}, 451 (1991).}, and most recently
Dine and Thomas \DT.

	Now consider the effect on this rate of a hypercharge
potential $\phi_Y$ in a small region of space.
The free energy density now includes the term
\eqn\ec{\eqalign{n_i y_i \phi_Y}}
where $y_i$ is the hypercharge of species i. We can easily see
then from $\eb$ that the free energy of the initial relative to the
final state
in a $B$ violating process
is unaffected since for such a process
\eqn\ed{\eqalign{\Sigma_i y_i \nu_i = 0  }}
so that no biasing of $B$ violation occurs. Equation \ed\ is
just hypercharge conservation, and the fact that
a hypercharge potential $\phi_Y$, or any other gauge charge
potential does not bias anomalous processes
is a result of the fact that these charges are anomaly free.
This point has also been  made in \DT.
The same conclusion applies to an external potential
for fermionic hypercharge $Y_f$ -- of itself it does nothing to
bias the electroweak anomalous processes.

	So how can such a potential alter the abundance of the particle
species in such a way as to bias ${B}$ violation? From the
arguments above it follows that the rate of any process
is unaltered by a potential
for $Y$, since hypercharge is conserved by all the processes
which are turned on. Thus for such a potential the answer is clearly
that $B$ violation cannot be biased. The $n$ dependence of the result
calculated
directs us to what is happening. A potential for $Y_f$
does bias processes which do not conserve $Y_f$, such as
those involving Higgs particles shown in Figure 1, which are taken to be
`fast'.
The
local thermal equilibrium distribution attained in the
presence of a potential $\phi_{Y_f}(x)$ is just
\eqn\ee{\eqalign{ f_i(x,p,t)
={{1}\over{e^{\beta(E_i+\mu_i)}\pm 1}} \quad;
              \quad E_i=|\vec{k}|+y_i^f\phi_{Y_f}(x). }}
This is an exact solution of the Boltzmann
equation for any static potential
$\phi_{Y_f}(x)$, in which
all collision terms
are identically zero, provided the usual relations
between the chemical potentials for the species involved in
a reaction are satisfied\footnote{$^\dagger$}{The expression
given for $E$ is more complicated when the Higgs vev is included.
The chiral coupling of hypercharge means that the correct dispersion
relation is not simply $\sqrt{k^2+m^2} +
y_i^f\phi_{Y_f}(x)$ as was stated in a preprint version
of this paper.}.
In the absence of any  constraints all the $\mu_i$
are zero and one concludes  that
there is no biasing of $B$ violation.

This brings us then to the crux of the matter:
what are the correct physical constraints on the particle
densities in a local region of space?

	First we note that it does
make sense to impose, by using an $x$-dependent
chemical potential, the constraint
that the {\it gauge} charge be zero everywhere. We do this
not because gauge charge is conserved
but in order to model the effects of screening: a local accumulation
of gauged charge will drag in an exactly cancelling charge from
the surrounding plasma. But as we have pointed out, the screening
is irrelevant to $B$ violation, because chemical potentials for charge $Q$ or
hypercharge $Y$ have no biasing effect on the anomalous processes.
The equilibrium distribution in a bubble wall is then
described by equation \ee\ with $\mu_i = q_i \mu_Q(x)$
in the broken phase, and $\mu_i = y_i \mu_Y(x)$ in the unbroken phase,
where $\mu_Q(x)$ and $\mu_Y(x)$ are determined by the conditions
that the net gauge charge density be everywhere zero
\ref\foot{ In the
unbroken phase one can constrain $Y=0$, and then
the nonabelian charges  like $T_3$ are  automatically zero by
symmetry. Alternatively one can constrain both $Q$ and $T_3$
to be zero. In the broken phase one should constrain $Q=0$.}.

The question of global quantum numbers requires more careful
consideration.  The
effect of imposing  global constraints like those enforced by
CKN is to force certain
linear combinations of the particle chemical potentials
$\mu_i$ non-zero. But is it physical to impose such constraints?
The region we are considering is surrounded
by an effectively infinite bath of global charge which is pulled
in by the applied or induced potentials. By imposing constraints
such as $B-L=0$ and $B=0$ one
is simply preventing some linear combination of particle species
from moving in response to these potentials in order to reach the
equilibrium given in \ee.

To resolve the question we need to consider the rate at which
the equilibrium distribution \ee\
is established in the bubble wall. To do this one turns
to the Boltzmann equation.
In the case of slow, thick walls discussed by CKN, the particles
passing the wall should be accurately described as a fluid
(because the particle mean free path is much smaller than
 the wall thickness).
Following the usual derivation of the linear perturbation equations
for a relativistic fluid
from the Boltzmann equation in the rest frame of the wall which moves
with velocity $v_w$ one arrives at
\eqn\eea{\eqalign{
\dot{n} +  v' &=0 \cr
\dot{\delta} + {4\over 3} v' &=
           -\Gamma_{\delta}(\delta-\epsilon n)  \cr
\dot{v} + {1\over 4} \delta' + {3 n_o\over 4\rho_o} \bigl(
y^f
\phi_{Y_f}{'}
- y E_x \bigr) &=-\Gamma_v v. \cr
}}
Here $n$ and $\delta$ are the fractional number and
energy density perturbations respectively,
and
$v$ the velocity of the fluid describing the species.
$n_o$ and $\rho_o$ are the unperturbed number and energy density,
$E_x$ the hypercharge electric field, and
$\Gamma_{\delta}$ and $\Gamma_v$ are velocity and temperature fluctuation
damping terms calculated from the collision integrals for
gauge boson exchange processes and
$\Gamma_v={\epsilon \over 4}D^{-1}$ where $D$ is
 the diffusion constant
for the species. Finally, primes denote spatial derivatives,
dots time derivatives, and
$\epsilon= {9 \zeta(3)^2 \over 7 \zeta(2) \zeta(4)}$, with
$\zeta(n)$ the Riemann zeta function. A full treatment involves considering
a set of equations like (8) for each particle species, coupled together
by decay terms
\ref\JPTlongone{M. Joyce, T. Prokopec and
N. Turok, Princeton preprint PUPT-1495.}.

For slow walls, as $v_w \rightarrow 0$, the solution approaches
the thermal distribution \ee\ with $\mu_i=0$.
The equations also resolve what might appear at first a paradox: in
the equilibrium distribution \ee\ many global conserved
or approximately conserved quantum numbers are nonzero. Yet they
must be zero for the universe as a whole. The fluid equations
tell us what is going on. The buildup of particles in the potential
$\phi_{Y_f}$ describes a state of hydrostatic equlibrium, in which
the force from the potential is balanced by the pressure of the
particle fluid. As particles of a given species enter the wall, they
slow down or speed up accordingly as their density
increases or decreases. This `time delay' effect means that as the
stationary state is being set up, deficits and excesses of particles
or antiparticles are left behind it.

More quantitatively, we must determine whether the
constrained or unconstrained equilibrium gives the leading approximation
to the particle chemical potentials for typical estimated parameters.
A simple estimate is as follows. The distance a particle with diffusion
constant $D$ diffuses in a given direction in a time $t$ is $\sim \sqrt{Dt}$.
In order that the transport processes be negligible the distance
a particle diffuses in the time the wall takes to
pass should be much less than the wall thickness $L$ i.e.
\eqn\eec{\eqalign{
v_w  >>  {D \over L }.
}}
This simple estimate is confirmed by an analysis of the stationary
solutions to equation \eea (see
\JPTlongone).
We have calculated the diffusion constant $D$
for quarks in the plasma, and find it is
$\approx (8 \alpha_s^2 T)^{-1}
\sim 6 T^{-1}$.
For `slow' walls ($ v_w <0.1$), $L$ must be considerably larger
than perturbative estimates of typical wall thicknesses
\ref\bub{N. Turok, Phys. Rev. Lett. {\bf 68}, 1803 (1992);
M. Dine, R. Leigh, P. Huet, A. Linde and D. Linde,
Phys. Rev. {\bf D46},550(1992);
B-H. Liu, L. McLerran and N. Turok, Phys. Rev. {\bf D46}, 2668 (1992).}.
(The mechanism applies to slowly moving  bubble walls
because the Higgs processes must  have time to
equilibrate as the wall passes i.e. $v_w<<(L\Gamma)^{-1}$, where $\Gamma$
is the rate of the Higgs process).

We conclude therefore that allowing the system to respond to
the potential induced on the bubble wall means that the equilibrium
calculated by CKN is not the appropriate one for the description
of typical `adiabatic' parameters. Beyond this we cannot
conclude what the baryon violation will be without a full calculation.
It is clear that the inclusion
of transport processes opens up the possibility of producing an asymmetry
in front of the wall where the $B$ violation is unsupressed
\ref\cknnew{This point
was made by M. Joyce, in the Proceedings
of the Sintra Conference on Electroweak Physics, May 1994,
and independently in
A. Cohen, D. Kaplan and A. Nelson, ``Diffusion Enhances Spontaneous
Baryogenesis'', preprint NSF-ITP-94-67, June 1994. }.
Here
we hope simply to  have clarified the fact that in setting
global quantum numbers zero one is assuming that
the conductivity of the plasma is zero, which is in fact
a very poor approximation.

We now turn to the `charge transport' mechanism. It is different in that
there are real physical constraints provided by the
asymmetry in particle fluxes reflected by the wall and
the physics is not simply that of a response to induced local
potentials for total or fermionic hypercharge.
The mechanism \cknnpb\ has a more involved history -- originally
CKN imposed similar
constraints on global quantum numbers, while they did {\it not}
impose any constraint on gauged hypercharge.  The latter
was corrected
in \cknplbb\ after Khlebnikov
\ref\khleb{S. Khlebnikov, Phys. Lett. {\bf B300}, 376 (1993).}
pointed out
that screening effects had been neglected. As we have
discussed screening effects do not alter the $B$ violation
rate. Indeed  CKN showed
that one could account for all screening
effects by going to a basis in which the global conserved
charges are `orthogonal' to hypercharge. These orthogonal charges
are simply the linear combinations which are unaffected by screening.
When one redoes their calculation in this new basis the result
is unchanged, since  in their corrected calculation  the reflected
hypercharge is still effectively  the quantity
which drives $B$ violation.

This does correctly account for screening if the global
quantum numbers taken to be zero locally are made non-zero by
screening processes only.
However once again it is incorrect to
set the various conserved and
approximately conserved quantum numbers to zero locally,
even without screening.
This becomes clear as soon as one thinks through the
microphysics. The wall reflects a chiral asymmetry in top quarks
(equal
excesses of left handed  anti-tops, and right handed tops).
These  asymmetries may be described by a chemical
potentials $\mu_{t_L}$ and $\mu_{t_R}$, the former of which
from equation \eb\ drives $B$ violation.
At the wall the reflected flux
carries none of each of the global charges set
zero by CKN. But particle interactions change this.
In particular, the Higgs/top processes are different for
the left and right handed particles
(see Figure 1).
In CKN's chemical potential calculation,
precisely these processes were assumed to be `fast'.
Note that the $\overline{t}_L$'s have a smaller interaction
rate than the $t_R$'s. Thus the $\overline{t}_L$ and
$t_R$ particle distributions are different
in front of the wall, and this causes
quantities like $B-L$ to be driven
nonzero locally. As in the case of screening,
particle interactions can
produce  nonzero densities of globally conserved charges.

If instead one assumes the Higgs/top quark interactions to
be `slow', then it is clear that hypercharge is not really relevant
to $B$ violation -- the driving force is simply the left handed
top quark asymmetry. The result one obtains differs from that of CKN
by a factor of $(2n+1)$, with $n$ being the number of Higgs doublets,
because in their calculation they
in effect share the
reflected hypercharge coming off the wall between
Higgs particles and top particles.

Since one cannot set globally conserved charges to zero locally,
it follows that one cannot argue that the driving force
for $B$ violation is simply hypercharge. But besides making
the calculation of the $B$ asymmetry apparently simple,
CKN had another reason for focusing on hypercharge,
 namely that
it is conserved in
all particle interactions and so
its propagation should be `more persistent'.
For example, in the processes in Figure 1, which are dominated by
forward scattering,
a ${t}_R$ can be converted into a $\phi^+$,
and this $\phi^+$ can later convert back into
a ${t}_R$ or a $\overline{b}_L$, resurrecting the chiral asymmetry
which drives $B$ violation.
This point has some merit,
 since Higgs particles
do not interact  strongly and so
propagate more efficiently than top quarks.

There {\it is} a correct, although more laborious
way to calculate the
$B$ asymmetry produced by an injected flux.
One should introduce local ($x$-dependent) chemical potentials
for each species which allow one to model the profiles of the
particle species involved.
One uses the appropriate Boltzmann  equations to
describe the propagation and decay of these potentials,
explicitly including the interactions responsible for the
decay of the chiral fermion asymmetry.

In a companion paper
\ref\jpt{M. Joyce, T. Prokopec and
N. Turok, Princeton preprint PUP-TH-1437 (1993).}
we have developed this technique for
the case of leptons, where each stage of the calculation from
quantum mechanical reflection through diffusion and
decay is possible in a well controlled approximation.
In this case the propagation of the injected flux very clearly
makes non-zero the local densities of various global charges
(in particular, lepton number $L$) because of the very different
diffusion properties of left- and right-handed leptons.
In that paper we argue that for typical  bubble walls
predicted by perturbative calculations \bub\
in the standard model and its minimal extension, the two Higgs theory,
 the
`lepton transport' mechanism can produce a considerable asymmetry
in this regime.

For the regime in which a fluid description is more appropriate, i.e.
for quarks at typical wall thicknesses, a full new treatment is required
using the fluid equations which we sketched above. Several other questions
remain to be resolved about the mechanism - the objections raised in
\gs\ and \DT\  must be addressed. Our aim here  has been
to point out an error which has propagated widely
in the literature.

\centerline{\bf Acknowledgements}

We thank M. Shaposhnikov for stimulating discussions,
A. Cohen, D Kaplan and A. Nelson for helpful correspondence,
and P. Malde for comments on the original version of this paper.
M.J. is supported by a Charlotte Elizabeth
Procter Fellowship.
The work of N.T. was partially	supported by
NSF contract PHY90-21984, and the David and Lucile Packard
Foundation.

\listrefs
\bye